\documentclass[twocolumn,pre,twoside,showpacs,preprintnumbers,floatfix]{revtex4}

\usepackage{amsmath}
\usepackage{amssymb}
\usepackage{graphicx}
\usepackage{dcolumn}
\usepackage{bm}

\def\d{\mathrm{d}}
\def\epsilon{\varepsilon}
\def\theta{\vartheta}
\def\rho{\varrho}
\def\vec#1{\mathbf{#1}}

\begin{document}


\title{Bulk and interfacial properties of binary hard-platelet fluids}

\author{M. Bier}
\email{bier@fluids.mpi-stuttgart.mpg.de}
\author{L. Harnau}
\author{S. Dietrich}

\affiliation{
   Max-Planck-Institut f\"ur Metallforschung,  
   Heisenbergstra\ss e 3, 
   D-70569 Stuttgart, 
   Germany,
}

\affiliation{
   Institut f\"ur Theoretische und Angewandte Physik, 
   Universit\"at Stuttgart, 
   Pfaffenwaldring 57, 
   D-70569 Stuttgart, 
   Germany
}

\date{October 30, 2003}

\begin{abstract}
   Interfaces between demixed fluid phases of binary mixtures of hard platelets are 
   investigated using density-functional theory. The corresponding excess free energy 
   functional is calculated within a fundamental measure theory adapted to the Zwanzig 
   model, in which the orientations of the particles of rectangular shape are restricted 
   to three orthogonal orientations. Density and orientational order parameter profiles 
   at interfaces between coexisting phases as well as the interfacial tension are 
   determined. A density inversion, oscillatory density profiles, and a Fisher-Widom line have 
   been found in a mixture of large thin and small thick platelets. The lowest interfacial 
   tension corresponds to the mean bulk orientation of the platelets being parallel to the interface. 
   For a mixture of large and small thin platelets, complete wetting of an isotropic-nematic interface 
   by a second nematic phase is found. 
\end{abstract}

\pacs{61.20.Gy, 61.30.Hn, 68.05.-n, 82.70.Dd}

\maketitle


\section{Introduction}

There is growing interest in thermal and structural properties of 
suspensions of platelike colloids like blood, clay sols, or liquid crystal dispersions because
of possible applications in biomedicine (e.g. shape-selective separation of cell components \cite{maso:02}),
geophysics (e.g. oil drilling \cite{mait:00}), or liquid crystal display technology \cite{bush:02}.
From theoretical considerations \cite{onsa:42,onsa:49} and computer simulations \cite{eppe:84,veer:92,case:95} 
one expects an isotropic to nematic transition due to the orientational degrees of freedom for sufficiently 
high platelet concentrations. This transition is not observed for clays because the long-ranged Coulomb 
interaction between the charged clay particles leads to gelation \cite{mour:95,gabr:96}. Nonetheless, recent 
preparation methods have been developed to produce suspensions of sterically stabilized platelets which do 
exhibit the expected isotropic to nematic transition \cite{brow:98,kooi:98,kooi:01,beek:03}. 
It was shown experimentally \cite{kooi:01} and theoretically 
\cite{bate:99,wens:01,harn:02a,mart:03} that polydispersity in the 
size of the platelets strongly affects the phase behavior. In particular, 
binary mixtures of thin and thick platelets lead to an unexpected 
isotropic-nematic density inversion \cite{kooi:01,wens:01}. 
Although attention has been paid to the bulk phase behavior \cite{harn:02a,mart:03}, so far there are no 
studies on fluid-fluid interfacial properties of such mixtures. On the basis of recent 
theoretical studies on fluids of thin hard platelets near hard walls \cite{harn:02b,harn:02c}, 
we expect that the competition of orientational entropy and excluded volume interaction leads to 
interesting fluid-fluid interfacial properties which should be experimentally accessible via optical 
techniques.

In this paper, we study interfacial properties between coexisting 
isotropic and nematic phases of a binary mixture of hard square cuboids within the
Zwanzig approximation \cite{zwan:63}. This model is chosen as the simplest non-trivial
approximation to the above-mentioned properties of real platelet suspensions. Binary mixtures are used to mimic polydispersity.
The hard-particle approximation is doubtful for long-ranged platelet-platelet interactions, but there are 
model systems of quasi-hard platelets \cite{brow:98,kooi:98}. The actual shape of the platelets (disks, stripes, hexagons, or 
square cuboids) is expected to be less important, only the thickness $L$ and the width $D$ are assumed to be relevant. Finally,
the square face of the platelets can take only three rather than a continuous range of orientations (Zwanzig model \cite{zwan:63}).
This model offers the advantage that the excess free energy functional can be determined within a fundamental measure theory 
\cite{cues:97a,cues:97b,cues:99,mart:99} and that the difficult determination of inhomogeneous density profiles becomes numerically 
feasible allowing one to study interfacial properties of binary hard-platelet fluids in detail. Due to the approximations, we only 
expect to find qualitatively correct results.

This paper is organized as follows. In Sec. \ref{sec2} we describe the density functional and the fundamental 
measure theory. Sec. \ref{sec3} presents two representative bulk phase diagrams 
of binary hard-platelet mixtures involving isotropic, nematic, and columnar
phases. In Sec. \ref{sec4} we determine the density and the orientational order parameter
profiles as well as the interfacial tensions of isotropic-nematic interfaces
in a mixture of large thin and small thick platelets. Sec. \ref{sec5} presents density profiles at 
nematic-nematic interfaces. The isotropic-nematic interfaces 
near an isotropic-nematic-nematic triple point are investigated in Sec. \ref{sec6}. 
Our results are summarized in Sec. \ref{sec7}.

\vfill


\section{\label{sec2}Density functional and fundamental measure theory}

We consider a binary mixture of hard rectangular particles of size 
$L_i \times D_i \times D_i$, $i \in \{1,2\}$. The position of the center of mass 
$\vec{r}$ is a continuous variable, whereas the normal of the square face is restricted 
to directions $\beta \in \{x,y,z\}$. The number density of the centers of mass of the 
platelets of size $i$ and orientation $\beta$ at position $\vec{r}$ is denoted by 
$\rho_{i,\beta}(\vec{r})$. In the absence of external potentials, the equilibrium density 
profiles of the mixture minimize the grand potential functional
\begin{eqnarray}
   \Omega[\{\rho_{i,\beta}\}] 
   & = & 
   \sum_{i,\beta}\int\d^3r\;
   \rho_{i,\beta}(\vec{r})\left(k_BT\left(\ln(\rho_{i,\beta}(\vec{r})\Lambda_i^3)-1\right)\right. \nonumber\\
   &   &
   \left.\vphantom{k_BT\left(\ln(\rho_{i,\beta}(\vec{r})\Lambda_i^3)-1\right)}-\mu_i\right) 
   + F^{\mathrm{ex}}[\{\rho_{i,\beta}\}],
   \label{eq1}
\end{eqnarray}
where the chemical potential $\mu_i$ and the thermal de Broglie wavelength $\Lambda_i$ of 
platelets of size $i$ are independent of the particle orientation.

In fundamental measure theory, one postulates the following form of the excess free energy functional 
\cite{cues:97a,cues:97b,cues:99,mart:99}:
\begin{equation}
   F^{\mathrm{ex}}[\{\rho_{i,\beta}\}] = k_BT \int \d^3r\;\Phi(\{n_l(\vec{r})\})
   \label{eq2}
\end{equation}
with the reduced excess free energy density
\begin{eqnarray}
   \Phi(\{n_l\}) & = & -n_0\ln(1-n_3) + \sum_{\sigma\in \{x,y,z\}}\frac{n_{1\sigma}n_{2\sigma}}{1-n_3} \\
                 &   & + \frac{n_{2x}n_{2y}n_{2z}}{(1-n_3)^2}
    \label{eq3}
\end{eqnarray}
and the weighted densities
\begin{equation} 
   n_l(\vec{r}) = \sum_{i,\beta}\int\d^3r'\;
   \omega^{(l)}_{i,\beta}(\vec{r}-\vec{r'})\rho_{i,\beta}(\vec{r'})
   \label{eq4}
\end{equation}
for $l \in \{0, 1x, 1y, 1z, 2x, 2y, 2z, 3\}$. 

The weight functions $\omega^{(l)}_{i,\beta}$ are obtained by expressing the Fourier 
transform of the Mayer-$f$ function as a sum of products of single particle functions. 
The Mayer-$f$ function equals $-1$ if the particles overlap and is zero otherwise. Explicit 
expressions of the weight functions are documented in Refs. \cite{cues:97a,cues:97b,cues:99,mart:99}. In the limit 
of infinitely thin platelets ($L_i/D_i\to 0$) the excess free energy density reduces to a 
third-order virial approximation \cite{harn:02c}. For convenience, we introduce 
the dimensionless variables $\mu^*_i := \mu_i/k_BT + 3\ln(D_1/\Lambda_i)$.


\section{\label{sec3}Bulk phase diagrams}

The bulk phase behavior of spatially homogeneous  binary platelet fluids
has recently been studied within the Zwanzig model \cite{harn:02a}. Rich 
phase diagrams are found involving an isotropic and one or two nematic phases 
characterized by different concentrations. The phase diagrams are very sensitive 
to the size ratio $D_1 / D_2$ and to the aspect ratios $L_1 / D_1$ and $L_2 / D_2$.

\begin{figure}[!bht]
   \includegraphics{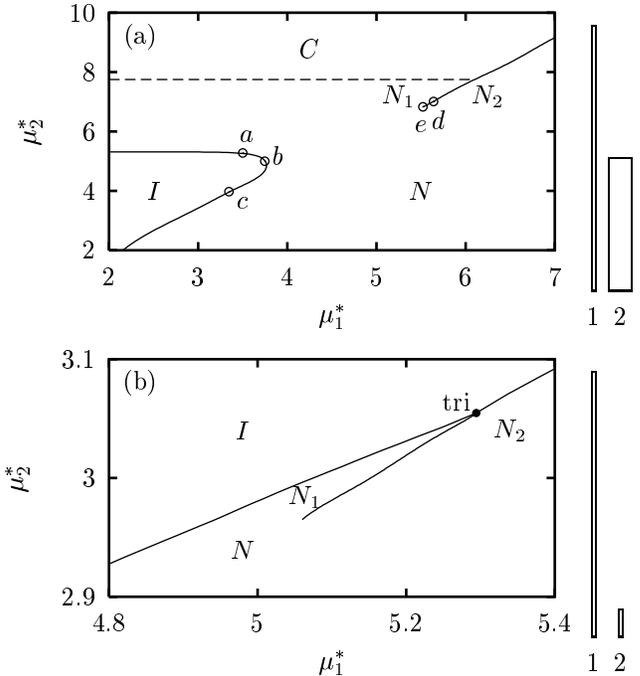}
   \caption{\label{fig1}Bulk phase diagrams of binary mixtures of large and small hard platelets ($L_1/D_1 = 0.01$; 
           (a): $D_1/D_2 = 2$, $L_2/D_2 = 0.16$; (b): $D_1/D_2 = 10$, $L_2/D_2 = 0.1$; see also the schematic side
           views $L_i \times D_i$ on the right side bar) in terms of the reduced chemical potentials $\mu^*_1$ and $\mu^*_2$. 
           Full lines represent strongly first-order transitions, whereas dashed lines denote second-order or weakly first-order
           transitions.
           In (a), one isotropic phase $I$, one nematic phase $N$, and one columnar phase $C$ are found. The isotropic-nematic
           transitions in the monodisperse limits occur at $\mu^*_1 = 0.829(1)$ for the large platelets 
           ($\mu^*_2 \rightarrow -\infty$) and at $\mu^*_2 = 5.505(1)$ for the small platelets ($\mu^*_1 \rightarrow -\infty$).
           Density and order parameter profiles for isotropic-nematic interfaces at state points $a$, $b$, and $c$ are shown
           in Fig. \ref{fig2}. The nematic phase $N$ demixes into two nematic phases $N_1$ and $N_2$. Fig. \ref{fig4} exhibits
           density profiles for nematic-nematic interfaces at state points $d$ and $e$, close to the lower critical point of the 
           $N_1$-$N_2$ demixing. For small values of $\mu^*_1$, the nematic-columnar transition is of second or weakly first order
           (dashed line), whereas it is strongly first-order (solid line) if $\mu^*_1$ is sufficiently large. The dashed line appears 
           to be straight because for this range of the chemical potentials, the density of the large platelets is negligible 
           compared to that of the small platelets.
           In (b), only a small part of the phase diagram around the triple point $\mathrm{tri}$ is shown,
           where the isotropic phase $I$ coexists with two nematic phases $N_1$ and $N_2$.} 
\end{figure}

Fig. \ref{fig1} 
shows two representative examples of bulk phase diagrams of binary mixtures 
of large and small platelets in terms of the reduced chemical potentials $\mu^*_1$ and $\mu^*_2$. 
The coexistence curves in these diagrams are calculated numerically by finding pairs of 
equilibrium states with equal chemical potentials and pressures. The phase transitions
at full curves are strongly first-order, whereas dashed lines denote second-order or weakly
first-order transitions. 

For a binary fluid consisting of large thin and small thick platelets (see Fig. \ref{fig1}(a)), 
a single nematic phase $N$ is found, which is separated from the isotropic phase $I$ by a first-order
phase boundary line. At larger values of the two chemical potentials, the nematic phase $N$ demixes 
discontinuously into two nematic phases $N_1$ and $N_2$. Furthermore, the nematic phase $N$ undergoes
a phase transition into a columnar phase $C$, where parallel columns of small platelets are surrounded by single 
large platelets with orientations perpendicular to the column axes. For small values of $\mu^*_1$, the nematic-columnar 
transition is of second or weakly first order. The phase boundary line appears straight because the density of the 
large platelets is negligible there, so that the mixture is nearly monodisperse. The nematic-columnar boundary line
changes to strongly first-order at the tricritical point where the nematic-nematic boundary line ends.

Upon decreasing the size of the small platelets, the lower critical point of this first-order $N_1$-$N_2$ 
coexistence curve shifts to smaller values of the chemical potentials until the $N_1$-$N_2$ and the 
$I$-$N$ coexistence curves start to intersect, giving rise to a triple point at which two nematic phases 
$N_1$ and $N_2$ coexist with the isotropic phase $I$ (see Fig. \ref{fig1}(b)). 


\section{\label{sec4}Isotropic-nematic interface}

In this section we discuss the properties of interfaces between the coexisting isotropic and nematic phases 
depicted in the phase diagram of Fig. \ref{fig1}(a). The interface normal is always arranged to point in $z$-direction.
The results are expressed in terms of the orientationally averaged number density
\begin{equation} 
   \rho_i(z) := \rho_{i,x}(z) + \rho_{i,y}(z) + \rho_{i,z}(z),
   \label{eq5}
\end{equation}
the nematic order parameter
\begin{equation} 
   s_i(z) := \frac{\rho_{i,z}(z) - \frac{1}{2}(\rho_{i,x}(z) + \rho_{i,y}(z))}{\rho_i(z)},
   \label{eq6}
\end{equation}
and the biaxial order parameter
\begin{equation}
   q_i(z) := \frac{\rho_{i,x}(z) - \rho_{i,y}(z)}{\rho_i(z)}.
\end{equation}

\begin{figure}[!bht]
   \includegraphics{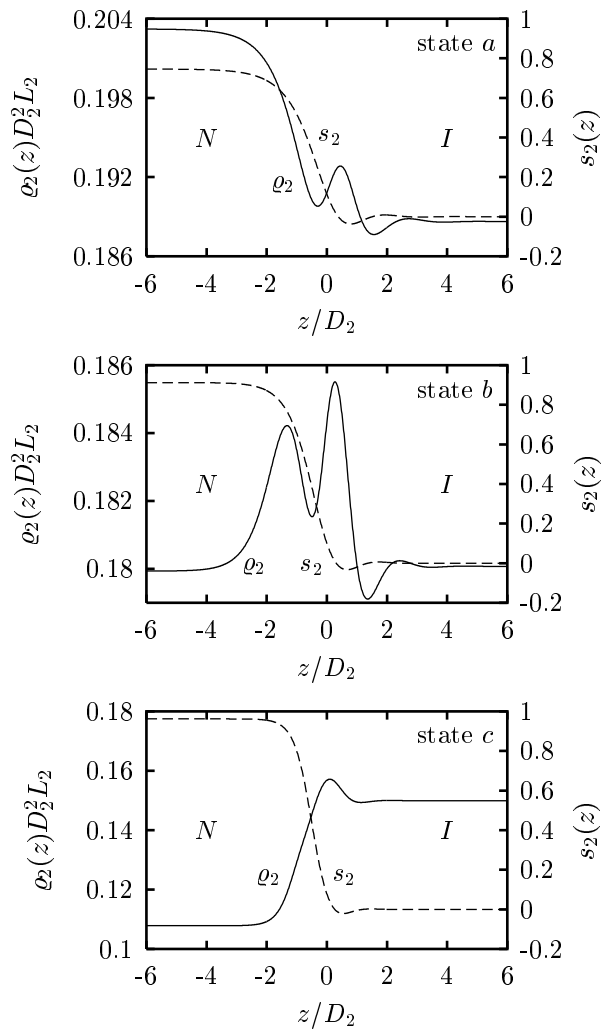}
   \caption{\label{fig2}Orientationally averaged density profiles $\rho_2$ (solid lines)
           and nematic order parameter profiles $s_2$ (dashed lines) of the small thick platelets 
           for the state points $a$, $b$, and $c$ in Fig. \ref{fig1}(a) at a planar interface between 
           the coexisting isotropic phase $I$ at $z\to\infty$ and the nematic phase $N$ at $z\to -\infty$.
           The position $z = 0$ of the origin is chosen to be given by $s_2(0) = \frac{1}{2}$. The interface normal and 
           the director of the nematic phase are parallel, which corresponds to the lowest surface tension
           (see Fig. \ref{fig3}). A density inversion occurs near the state point $b$. At the interface layers form which
           slightly prefer platelet orientations perpendicular to the interface normal.}
\end{figure}

Fig. \ref{fig2}
shows the density and the nematic order parameter profiles of the small thick platelets 
for the state points denoted as $a$, $b$, and $c$ in Fig. \ref{fig1}(a). The interface normal and the director of the
nematic phase are parallel and the position $z = 0$ of the origin is fixed by choosing $s_1(0) = \frac{1}{2}$. 
No biaxiality has been observed, that is, $q_i(z) \equiv 0$.
We have confirmed that the density profiles for both thin and thick platelets show oscillations of the same period,
but the amplitude for the thin platelets is considerably smaller than for the thick platelets. 

Whereas the density of the thin platelets is always smaller in the isotropic phase than in the
nematic phase along the coexistence curve, a density inversion for the thick platelets is found.
For state $b$, the density of the small thick platelets in the isotropic 
and the nematic phase are nearly identical. This remarkable phenomenon of 
isotropic-nematic density inversion has been observed experimentally 
for a dispersion of sterically stabilized gibbsite ($\mathrm{Al(OH)_3}$) platelets \cite{kooi:01}.

For isotropic-anisotropic coexistence in binary mixtures of long and short thin rods, one observes a large fraction of long 
rods in the anisotropic phase if there is a small fraction of the long rods in the isotropic phase \cite{lekk:84}. For platelets this
fractionation does not occur to the same extend because the density $\rho_1$ increases by at most one order of 
magnitude upon traversing the transition from the isotropic to the nematic phase.

The nematic order parameter profiles shown in Fig. \ref{fig2}  exhibit 
a small negative minimum ($s_2(z) < 0$) on the isotropic side of the interface
implying a depletion of platelets with orientation parallel to the interface. A similar
minimum has been found for freely rotating 
platelets near a hard wall \cite{harn:02b}. This indicates that the 
nematic phase acts like a wall creating a corresponding depletion zone in the
adjacent isotropic phase. 

\begin{figure}[!bht]
   \includegraphics{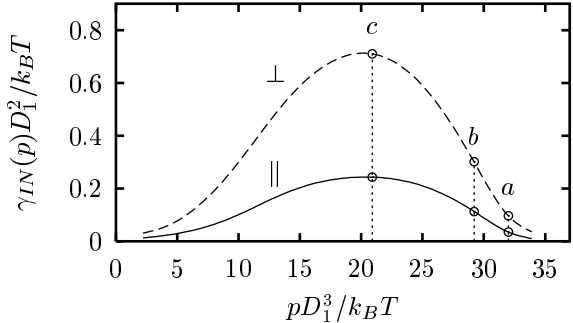}
   \caption{\label{fig3}Isotropic-nematic interfacial tension $\gamma_{IN}$ of a binary mixture of large thin and small 
           thick platelets (see Fig. \ref{fig1}(a)) as a function of the pressure $p$ along the line of phase coexistence.
           Through the choice of the boundary condition for the orientation of the nematic order at $z \rightarrow -\infty$ the 
           director of the nematic phase can be parallel ($\|$) or perpendicular ($\perp$) to the interface normal. 
           The monodisperse fluids, consisting of either the large or the small platelets, correspond to the endpoints of 
           the curves at the left and the right, respectively. The density profiles and the orientational order 
           parameter profiles for the thermodynamic states marked $a$, $b$ and $c$ are shown in  Fig. \ref{fig2}.}
\end{figure}

The surface tension $\gamma_{IN}$ of an isotropic-nematic interface with the director of the nematic phase at $z \rightarrow -\infty$
parallel ($\|$) or perpendicular ($\perp$) to the interface normal is shown in Fig. \ref{fig3} 
as a function of the pressure $p$ along the line of coexisting phases. 
For all pressures, the parallel configuration has the lower surface tension by a factor of rather precisely three.
As is apparent from Fig. \ref{fig3}, the surface tension is a non-monotonic function of the pressure and 
exhibits a maximum close to state $c$. The maximum value of $\gamma_{IN}$ is about $20$ times larger 
than the ones for the monodisperse fluids corresponding to the endpoints of the curves. This implies that the interface 
of the \emph{binary} fluid is rather stiff as compared to the ones of the corresponding monodisperse fluids. 

Calculating the typical surface tension at the inversion point $b$ ($\gamma^{\|}_{IN}D_1^2/k_BT \approx 0.1$) for the above
mentioned sterically stabilized gibbsite platelets ($D_1 \approx 170\mathrm{nm}$) at room temperature $T = 300\mathrm{K}$
yields $\gamma^{\|}_{IN} \approx 14\mathrm{nN/m}$, which perfectly agrees with the estimate provided in Ref. \cite{kooi:01}.

A similar behavior has been found for binary mixtures of thin and thick rods \cite{shun:01}. 


\section{\label{sec5}Nematic-nematic interfaces} 

\begin{figure}[!bht]
   \includegraphics{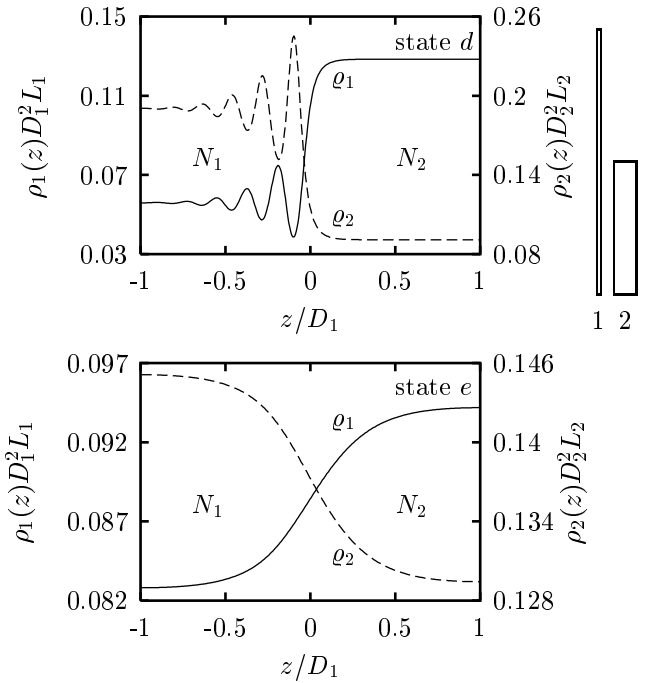}
   \caption{\label{fig4}The density profiles $\rho_1$ (solid lines)
           and $\rho_2$ (dashed lines) of the large thin and small thick platelets 
           of the binary hard-platelet mixture, respectively, at the planar interface between 
           two coexisting nematic phases ($N_1$ and $N_2$) for the states $d$ and $e$ indicated in 
           Fig. \ref{fig1}(a).}
\end{figure}

We now turn our attention to the properties of interfaces between two 
coexisting nematic phases whose corresponding phase boundaries are depicted in the phase diagram 
of Fig. \ref{fig1}(a). Fig. \ref{fig4}
shows the density profiles of the large and small platelets for such interfaces.
Pronounced oscillations are found for the state $d$ (see Fig. \ref{fig1}(a)) on that side of the interface 
which is rich in the small thick platelets. The wave length of the oscillations 
$\lambda = 0.175(5)D_1$ is more than twice the thickness of the thick platelets. 
The exponential decay length $\xi = 0.131(4)D_1$ is of the same order of magnitude. The
wave length $\lambda$ and the exponential decay length $\xi$ are the same for the large and the small platelets. 

For the state $e$, closer to the lower critical point of nematic-nematic coexistence, there is no sign of oscillations on 
either side of the interface. Upon approaching the lower critical point along the nematic-nematic coexistence curve, the 
density profiles $\rho_1$ and $\rho_2$ turn into monotonic functions of $z$ and the interfacial width broadens.
In simple fluids, the disappearance of oscillations in the density profiles is connected to the existence 
of the so-called Fisher-Widom line \cite{fish:69}, which divides the bulk phase diagram into regions where the 
asymptotic decay of bulk two-point correlation function is either monotonic or exponentially damped oscillatory \cite{evan:93,evan:94};
this behavior of the two-point correlation function carries over to the density profiles. Recently, such Fisher-Widom lines have 
been found for a binary Gaussian core model \cite{arch:01} and a model colloid-polymer mixture \cite{brad:02}.
For the binary platelet model under consideration, the determination of the Fisher-Widom line requires the solution 
of Ornstein-Zernike equations for a six-component system, two species and three allowed orientations, which is 
beyond the scope of the present study. However, from the observed vanishing of the oscillations of the density profiles 
we conclude that there is a Fisher-Widom line and it intersects the coexistence curve between $d$ and $e$.

We have also calculated the surface tensions of nematic-nematic interfaces for different relative orientations of the platelets
and the interface. Again the smallest value is taken by the configuration with the mean bulk orientation of the platelets being 
parallel to the interface (compare Fig. \ref{fig3}). Upon approaching the lower critical point of nematic-nematic coexistence, the 
expected mean-field-like vanishing of the surface tension and of the $N_1$-$N_2$ density difference has been recovered.


\section{\label{sec6}Isotropic-nematic interface near a triple point}

In this section we consider interfacial properties of coexisting isotropic 
and nematic phases near the triple point in the bulk phase 
diagram of Fig. \ref{fig1}(b). We choose $\epsilon := \mu^*_2 - \mu^{*\mathrm{tri}}_2 > 0$ 
at $I$-$N_2$ coexistence as a measure for the undersaturation with respect to the chemical potential 
of the small platelets at the triple point $\mu^{*\mathrm{tri}}_1 = 5.294376743251(1)$,
$\mu^{*\mathrm{tri}}_2 = 3.054579122021(1)$. Notice that for the present Zwanzig model the accuracy 
of the bulk coexistence data is higher than those one would be able to achieve for models of continuously rotating
platelets because only a finite set of coexistence equations has to 
be solved instead of nonlinear integral equations. Hence, the 
undersaturation $\epsilon$ can be defined rather precisely in the limit 
$\epsilon \rightarrow 0$ within the present model. This is crucial for the subsequent wetting analysis and motivates
the use of the Zwanzig approximation. 

\begin{figure}[!bht]
   \includegraphics{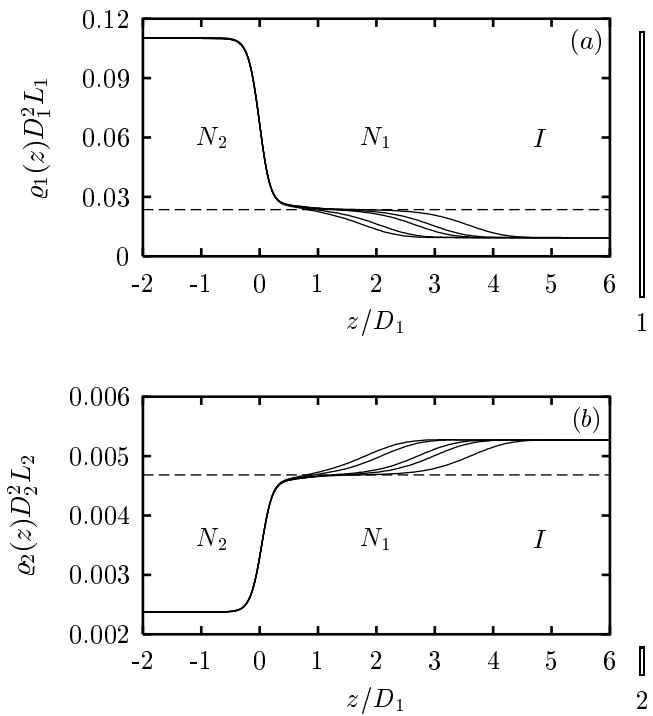}
   \caption{\label{fig5}Equilibrium density profiles of large (a) and small (b)
           thin platelets (see Fig. \ref{fig1}(b)) at various triple point undersaturations 
           $\epsilon := \mu^*_2-\mu^{*\mathrm{tri}}_2 \in \{10^{-3}, 5 \times 10^{-4},
           10^{-4}, 5 \times 10^{-5}, 10^{-5}\}$ (from left to right) at $I$-$N_2$ coexistence. The position $z=0$
           of the origin is chosen as the midpoint of the bulk densities of $N_1$ and $N_2$. The dashed 
           lines represent the densities of the nematic phase $N_1$ at the triple point.
           The isotropic phase $I$ is at $z \rightarrow \infty$, and the nematic phase $N_2$ is 
           at $z \rightarrow -\infty$.}
\end{figure}

Fig. \ref{fig5}
shows the density profiles of the large and the small platelets for various values 
of $\epsilon$. The density profiles exhibit a plateau region, signaling the onset of the formation of the still
metastable $N_1$ phase. In the limit of $\epsilon \rightarrow 0$, the local density of the plateau region agrees with the one
of the bulk phase $N_1$ at the triple point, whereas the asymptotic densities 
at $z \rightarrow\pm\infty$ are the ones of the coexisting phases $I$ and $N_2$.

\begin{figure}[!bht]
   \includegraphics{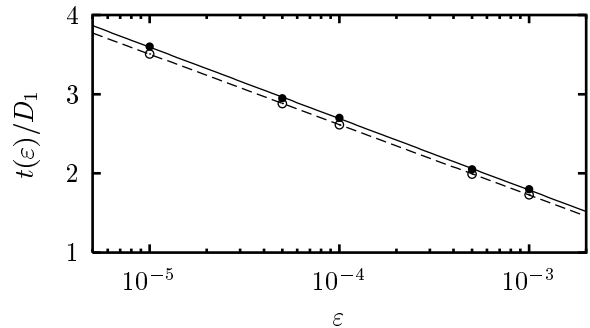}
   \caption{\label{fig6}Thickness $t$ of the $I$-$N_2$ interface as a function of the undersaturation 
           $\epsilon := \mu^*_2 - \mu^{*\mathrm{tri}}_2$ from the triple point (see Figs. \ref{fig1}(b) and \ref{fig5}). 
           The solid and the dashed line are calculated according to the inflection point and the midpoint method, respectively,  
           as described in the text. The correlation length $\xi=0.389(3)D_1$ is determined from the slope of the logarithmic 
           growth of $t$ for $\epsilon \rightarrow 0$.}
\end{figure}

Fig. \ref{fig6} 
shows the interface thickness $t$ as a function of the
undersaturation $\epsilon$. There are various possible definitions for the interface thickness.
We have used the following two measures:
\begin{itemize}
   \item The inflection point method: The interface thickness is given by 
         the distance between the outermost inflection points of the density 
         profiles.

   \item The midpoint method: The distance between the midpoints of the density profiles for the 
         $I$-$N_1$ and the $N_1$-$N_2$ interface defines the interfacial thickness.
\end{itemize}
Upon approaching the triple point, we find that $t$ diverges for $\epsilon \rightarrow 0$ as $t \simeq -\xi\ln\epsilon + t_1$ 
with $\xi = 0.392(4)D_1$, $t_1 = -0.92(4)D_1$ for the inflection point method and $\xi = 0.386(2)D_1$, $t_1 = -0.94(2)D_1$ for the 
midpoint method (see Fig. \ref{fig6}). The logarithmic divergence of $t$ is consistent with complete wetting of the 
$I$-$N_2$ interface by a $N_1$ film at the triple point in the absence of algebraically decaying interaction potentials 
\cite{diet:88}. 
Moreover, we have confirmed that the value of $\xi$ determined from the thickness of the interface is in agreement with the 
correlation length of the wetting phase $N_1$ at the triple point. The latter has been obtained from the asymptotic exponential 
decay of the density profiles from $I$-$N_1$ or $N_1$-$N_2$ interfaces towards their $N_1$ bulk values at the triple point.

The importance of interface fluctuations is determined by the so-called wetting parameter \cite{schi:90}
\begin{equation}
   \omega = \frac{k_BT}{4\pi\gamma_{IN_2}\xi^2},
\end{equation}
which for the present system takes a value of order unity at the triple point. Fluctuations are therefore neither negligible nor 
dominant.

In order to validate complete triple point wetting by an alternative method, the vanishing of the dihedral angle 
$\theta_{N_1}$ of a lense of phase $N_1$ at the triple point has been checked using the relation
\begin{equation}
   \cos \theta_{N_1} =
   \frac{\left(\gamma_{IN_2}^{\mathrm{tri}}\right)^2 - \left(\gamma_{IN_1}^{\mathrm{tri}}\right)^2
         - \left(\gamma_{N_1N_2}^{\mathrm{tri}}\right)^2}
        {2\gamma_{IN_1}^{\mathrm{tri}}\gamma_{N_1N_2}^{\mathrm{tri}}}.
\end{equation}
The interfacial tensions $\gamma_{IN_1}^{\mathrm{tri}}=0.05003056658 k_BT/D_1^2$ and 
$\gamma_{N_1N_2}^{\mathrm{tri}}=0.05600483075 k_BT/D_1^2$ are obtained by approaching 
the triple point along the $I$-$N_1$ and the $N_1$-$N_2$ coexistence curves, respectively, from below in terms of 
the chemical potential of the small platelets $\mu^*_2$ (see Fig. \ref{fig1}(b)).

Recently, complete wetting of an $I$-$N_2$ interface by a $N_1$ film has been 
predicted for a binary mixture of long thin and thick rods ($L>D$) with $L_1 = L_2, D_1/D_2 \geq 4$ \cite{shun:02,shun:03}. 
The ratio $\xi / L_1$ in those systems is of the same order of magnitude as the value $\xi / D_1 = 0.389(3)$ we have obtained 
for a binary mixture of large and small platelets ($L<D$) with $L_1 = L_2, D_1/D_2 = 10$.

At the emerging $I$-$N_1$ interface, the orientational order parameter profiles $s_1$ and $s_2$ exhibit slightly negative
minima indicating the depletion of platelets with orientation parallel to the interface. The absolute values of these minima are 
two orders of magnitude smaller for $s_2$ than for $s_1$.


\section{\label{sec7}Summary}

We have studied bulk and interfacial properties of binary hard-platelet fluids 
using density-functional theory. The platelets form square 
parallelepipeds with orientations restricted to three mutually 
perpendicular directions. A fundamental measure theory is used to 
derive the excess free energy functional. The grand potential functional 
is minimized numerically and phase diagrams, density profiles, orientational 
order parameter profiles, and surface tensions are determined leading to the 
following main results:
\begin{enumerate}
   \item The bulk phase diagrams for two representative examples of 
         binary hard-platelet fluids (Fig. \ref{fig1}) involve 
         an isotropic and one or two nematic phases of different concentration
         as well as a columnar phase.

   \item For a binary platelet mixture of large thin and small thick
         platelets, the density profiles of thick platelets at the 
         isotropic-nematic interface exhibit oscillations (Fig. \ref{fig2}). 
         A density inversion for the thick platelets is found. Near the thermodynamic state marked as $b$ in 
         Fig. \ref{fig2}, the densities of the thick platelets in the isotropic and 
         the nematic phase are nearly identical. The interfacial tension for platelets with their mean bulk orientation
         parallel to the interface is smaller than for the corresponding perpendicular configuration. The bidisperse 
         mixtures have significantly larger interfacial tensions than the corresponding monodisperse fluids (Fig. \ref{fig3}).

   \item Both the large  platelet and the small platelet density profiles 
         exhibit pronounced oscillations on one side of the nematic-nematic 
         interface, provided the chemical potentials are sufficiently high
         (Fig. \ref{fig4}, state $d$). Upon approaching the lower critical point of nematic-nematic demixing
         the oscillations vanish and the interface broadens (Fig. \ref{fig4}, state $e$).

   \item Just above the isotropic-nematic-nematic triple point shown in  Fig. \ref{fig1}(b), the $I$-$N_2$ 
         interfacial profiles thicken due to the formation of the incipient nematic phase $N_1$ at this interface
         (Fig. \ref{fig5}). Fig. \ref{fig6} demonstrates that the thickness of the wetting film diverges logarithmically upon 
         approaching the triple point. Complete triple point wetting is confirmed explicitly by observing the vanishing 
         of the dihedral angle at the triple point.
\end{enumerate} 


\begin{acknowledgments}
   The authors thank R. van Roij for useful discussions.
\end{acknowledgments}




\begin{thebibliography}{00}
   \bibitem{maso:02} T.G. Mason, Phys. Rev. E {\bf 66}, 060402(R) (2002).
   
   \bibitem{mait:00} G.C. Maitland, Curr. Opin. Colloid Interface Sci. {\bf 5}, 301 (2000).

   \bibitem{bush:02} R.J. Bushby and O.R. Lozman, Curr. Opin. Colloid Interface Sci. {\bf 7}, 343 (2002).

   \bibitem{onsa:42} L. Onsager, Phys. Rev. {\bf 62}, 558 (1942).
 
   \bibitem{onsa:49} L. Onsager, Ann. N. Y. Acad. Sci. {\bf 51}, 627 (1949).

   \bibitem{eppe:84} R. Eppenga and D. Frenkel, Mol. Phys. {\bf 52}, 1303 (1984).

   \bibitem{veer:92} J.A.C. Veerman and D. Frenkel, Phys. Rev. A {\bf 45}, 5632 (1992).

   \bibitem{case:95} A. Casey, P. Harrowell, J. Chem. Phys. {\bf 103}, 6143 (1995).

   \bibitem{mour:95} A. Mourchid, A. Delville, J. Lambard, E. L\'ecolier, and P. Levitz, Langmuir {\bf 11}, 1942 (1995).
   
   \bibitem{gabr:96} J.-C.P. Gabriel, C. Sanchez, and P. Davidson, J. Phys. Chem. {\bf 100}, 11139 (1996).

   \bibitem{brow:98} A.B.D. Brown, S.M. Clarke, A.R. Rennie, Langmuir {\bf 14}, 3129 (1998).

   \bibitem{kooi:98} F.M. van der Kooij and H.N.W. Lekkerkerker, J. Phys. Chem. B {\bf 102}, 7829 (1998).

   \bibitem{kooi:01} F.M. van der Kooij, D. van der Beek, and H.N.W. Lekkerkerker, J. Phys. Chem. B {\bf 105}, 1696 (2001).

   \bibitem{beek:03} D. van der Beek and H.N.W. Lekkerkerker, Europhys. Lett. {\bf 61}, 702 (2003).

   \bibitem{bate:99} M.A. Bates and D. Frenkel, J. Chem. Phys. {\bf 110}, 6553 (1999).

   \bibitem{wens:01} H.H. Wensink, G.J. Vroege, and H.N.W. Lekkerkerker, J. Phys. Chem. B {\bf 105}, 10610, (2001).

   \bibitem{harn:02a} L. Harnau, D. Rowan, and J.-P. Hansen, J. Chem. Phys. {\bf 117}, 11359 (2002).

   \bibitem{mart:03} Y. Mart\'\i nez-Rat\'on and J.A. Cuesta, J. Chem. Phys. {\bf 118}, 10164 (2003).

   \bibitem{harn:02b} L. Harnau and S. Dietrich, Phys. Rev. E {\bf 65}, 021505 (2002).

   \bibitem{harn:02c} L. Harnau and S. Dietrich, Phys. Rev. E {\bf 66}, 051702 (2002).

   \bibitem{zwan:63} R. Zwanzig, J. Chem. Phys. {\bf 39}, 1714 (1963).

   \bibitem{cues:97a} J.A. Cuesta and Y. Mart\'\i nez-Rat\'on, Phys. Rev. Lett. {\bf 78}, 3681 (1997).

   \bibitem{cues:97b} J.A. Cuesta and Y. Mart\'\i nez-Rat\'on, J. Chem. Phys. {\bf 107}, 6379 (1997).

   \bibitem{cues:99} J.A. Cuesta and R.P. Sear, Eur. Phys. J. B {\bf 8}, 233 (1999).

   \bibitem{mart:99} Y. Mart\'\i nez-Rat\'on, J.A. Cuesta, R. van Roij, and B. Mulder, in {\it New approaches 
                     to problems in liquid state theory}, edited by C. Caccamo, J.-P. Hansen, and G. Stell 
                     (Kluwer Academic, Dordrecht, 1999), NATO science series Vol. C {\bf 529}, p. 139.

   \bibitem{lekk:84} H.N.W. Lekkerkerker, Ph. Coulon, R. van der Haegen, J. Chem. Phys. {\bf 80}, 3427 (1984).

   \bibitem{shun:01} K. Shundyak and R. van Roij, J. Phys.: Condens. Matter {\bf 13}, 4789 (2001).

   \bibitem{fish:69} M.E. Fisher and B. Widom, J. Chem. Phys. {\bf 50}, 3756 (1969).

   \bibitem{evan:93} R. Evans, J.R. Henderson, D.C. Hoyle, A.O. Parry, and Z.A. Sabeur, Mol. Phys. {\bf 80}, 755 (1993).

   \bibitem{evan:94} R. Evans, R.J.F. Leote de Carvalho, J.R. Henderson, and D.C. Hoyle, J. Chem. Phys. {\bf 100}, 591 (1994).

   \bibitem{arch:01} A.J. Archer and R. Evans, Phys. Rev. E {\bf 64}, 041501 (2001).

   \bibitem{brad:02} J.M. Brader, R. Evans, M. Schmidt, and H. L\"owen, J. Phys.: Condens. Matter {\bf 14}, L1 (2002).

   \bibitem{diet:88} S. Dietrich, in {\it Phase Transitions and Critical Phenomena}, edited by C. Domb and 
                     J.L. Lebowitz (Academic, London, 1988), Vol. 12, p. 1.

   \bibitem{schi:90} M. Schick, in {\it Les Houches, Session XLVIII, 1988 --- Liquides aux interfaces /
                     Liquids at interfaces}, edited by J. Charvolin, J.F. Joanny, and J. Zinn-Justin
                     (North-Holland, Amsterdam, 1990), p. 415.

   \bibitem{shun:02} K. Shundyak and R. van Roij, Phys. Rev. Lett. {\bf 88}, 205501 (2002).

   \bibitem{shun:03} K. Shundyak and R. van Roij, cond-mat/0309414 (2003).
\end{thebibliography}
\end{document}